\documentclass[prl,amsmath,amssymb,twocolumn]{revtex4-1}

\pdfoutput=1

\usepackage{graphicx}
\usepackage{subfigure}
\usepackage{adjustbox}
\usepackage{bm}
\usepackage{color}
\usepackage{braket}
\usepackage{standalone}
\usepackage{multirow}
\usepackage{tikz}
\usepackage{mathrsfs}
\usepackage{dsfont}
\usepackage[colorlinks,bookmarks=true,citecolor=blue,linkcolor=red,urlcolor=blue]{hyperref}
\usepackage{epstopdf}

\newcommand{\bea}{\begin{eqnarray}}
\newcommand{\eea}{\end{eqnarray}}
\usepackage{float}
\usepackage{array}
\usepackage{slashed,bbold}

\usepackage{amsmath,amssymb,bm,bbm}
\usepackage{graphicx}

\usepackage[papersize={8.5in,11in}]{geometry}

\definecolor{darkblue}{rgb}{0.,0.,0.4}
\definecolor{darkred}{rgb}{0.5,0.,0.}
\definecolor{BlueViolet}{RGB}{138,43,226}
\definecolor{SkyBlue}{RGB}{30,144,255}
\definecolor{DarkGreen}{RGB}{0,100,0}
\begin{document}

\title{Higher-order exceptional point and Landau-Zener Bloch oscillations\\ in driven non-Hermitian photonic Lieb lattices}

\author{Shiqiang Xia$^{1,2}$ Carlo Danieli$^{3}$ Yingying Zhang$^{1}$ Xingdong Zhao$^{1,5}$ Hai Lu$^{1}$ Liqin Tang$^{2,6}$ Denghui Li$^{2}$ Daohong Song$^{2}$, and Zhigang Chen$^{2,4}$}

\affiliation{$^1$Engineering Laboratory for Optoelectronic Technology and Advanced Manufacturing, School of Physics, Henan Normal University, Xinxiang, Henan 453007, China
$^2$The MOE Key Laboratory of Weak-Light Nonlinear Photonics, TEDA Applied Physics Institute and School of Physics, Nankai University, Tianjin 300457, China
$^3$Max Planck Institute for the Physics of Complex Systems, Nothnitzer Stra?e 38, 01187 Dresden, Germany
$^4$Department of Physics and Astronomy, San Francisco State University, San Francisco, California 94132, USA
$^5$phyzhxd@gmail.com
$^6$tanya@nankai.edu.cn}

\begin{abstract}
{We propose a scheme to realize parity-time (PT) symmetric photonic Lieb lattices of ribbon shape and complex couplings, thereby demonstrating the higher-order exceptional point (EP) and Landau-Zener Bloch (LZB) oscillations in presence of a refractive index gradient. Quite different from non-Hermitian flatband lattices with on-site gain/loss, which undergo thresholdless PT symmetry breaking, the spectrum for such quasi-one-dimensional Lieb lattices has completely real values when the index gradient is applied perpendicular to the ribbon, and a triply degenerated (third-order) EP with coalesced eigenvalues and eigenvectors emerges only when the amplitude of gain/loss ratio reaches a certain threshold value. When the index gradient is applied parallel to the ribbon, the LZB oscillations exhibit intriguing characteristics including asymmetric energy transition and pseudo-Hermitian propagation as the flatband is excited. Meanwhile, a secondary emission occurs each time when the oscillatory motion passes through the EP, leading to distinct energy distribution in the flatband when a dispersive band is excited. Such novel phenomena may appear in other non-Hermitian flatband systems. Our work may also bring insight and suggest a photonic platform to study the symmetry and topological characterization of higher-order EPs that may find unique applications in for example enhancing sensitivity.}
\end{abstract}

\maketitle

{\it Introduction.---} In the past decade, the physics of dispersionless flatbands has attracted considerable interest, partly due to their connection to various fascinating phenomena including the fractional quantum Hall effect \cite{PhysRevLett.106.236802,PhysRevLett.106.236803,PhysRevLett.106.236804}, high-temperature superconductivity \cite{PhysRevB.83.220503}, and tunable strictly compact nonlinear excitations \cite{PhysRevA.87.061803,PhysRevE.92.032912,PhysRevA.96.063838,DanieliLTP}, among many others. Flatband phenomena also emerge in twisted bilayer graphene lattices that serve as a new platform for exploration of correlated phases \cite{cao2018correlated,balents2020supercond,wang2020localization}. The flatband eigenmodes are typically compact localized states (CLSs)
in real space due to local destructive interferences of the Bloch wave functions. Such CLSs have been observed in various systems, particularly in artificial photonic lattices \cite{PhysRevLett.114.245503,PhysRevLett.114.245504,Xia:16,Zong16}. For instance, the first direct experimental observation of CLSs was realized in Lieb photonic lattices \cite{PhysRevLett.114.245503,PhysRevLett.114.245504}, which stimulated a surge of  theoretical and experimental studies devoted to flatband physics - for a survey of the state of the art, see \cite{Derzhko2015,Leykam2018,LeykamAPL,tang2020photonic,Vicencio}. Efforts in realizing synthetic magnetic flux for light have allowed to observe Aharonov-Bohm caging in photonic rhombic lattices, an effect that originates from flatbands \cite{PhysRevLett.121.075502,kremer2020square}. Recently, investigations on flatbands have been extended to parity-time (PT) symmetric non-Hermitian systems \cite{feng2017non,el2018non,ozdemir2019parity}. The existence of second-order non-Hermitian degeneracies and the higher-order exceptional points (EPs) have led to many intriguing features such as unidirectional invisibility, topological phase transitions, PT-symmetric lasers, and even enhanced sensitivity  \cite{PhysRevLett.106.213901,PhysRevLett.115.040402,xia2021nonlinear,Feng2014single,hodaei2014parity,hodaei2017,delplace2021,mandal2021}. It has been shown that flatbands with completely real eigenvalues can appear in many non-Hermitian systems. For example, a bipartite sublattice symmetry enforces the existence of non-Hermitian flatbands, which are typically embedded in an auxiliary dispersive band - akin to the flatband bound states in the continuum \cite{PhysRevB.96.064305}. Defect states can emerge from a non-Hermitian flatband of photonic zero modes \cite{PhysRevLett.120.093901}. In addition, flatbands can arise due to the interplay between synthetic magnetic flux and non-Hermiticity \cite{PhysRevA.96.011802,PhysRevA.99.033810}. Recently, non-Hermitian flatband lattices have been demonstrated in experiment using laser-written photonic waveguide arrays with tailored loss distributions \cite{PhysRevLett.123.183601}.

On the other hand, Bloch oscillations, the oscillatory motion of particles in a periodic potential driven by an external force, is one of the most striking wave dynamics in condensed matter physics \cite{bloch1928quantum}. Landau-Zener tunneling between Bloch bands, a counteracting effect limiting Bloch oscillations, appears at the edges of the Brillouin zone where the band gap is shallow \cite{zener1934theory}. While Bloch oscillations have been demonstrated in various wave systems including optical waves \cite{PhysRevLett.83.4752,PhysRevLett.83.4756,PhysRevLett.91.263902,corrielli2013fractional}, typical examples of Landau-Zener Bloch (LZB) oscillations are better represented by their classical analogue observed in photonic waveguide arrays \cite{PhysRevLett.96.053903,PhysRevLett.121.033904}. Moreover, flatband Bloch oscillations together with triply Landau-Zener tunneling between the flatband and dispersive bands have also been proposed in optical lattices \cite{PhysRevLett.116.245301,long2017topological,XiaAPL}. Quite different from the Hermitian systems, the complex phase in non-Hermitian lattices indicates non-conserved energy, resulting in super-Bloch oscillations with amplification or damping, and even chiral Zener tunneling at the non-Bloch-band collapse point \cite{PhysRevLett.103.123601,PhysRevLett.124.066602}. In PT-symmetric photonic lattices, noticeable secondary emissions have been observed due to the degeneracy of the EPs \cite{wimmer2015obser,xu2016experimental}. Very recently, Bloch oscillations and Landau-Zener tunneling have been investigated in a PT-symmetric flatband rhombic lattice by introducing on-site gain/loss \cite{PhysRevA.103.023721}. However, due to the band touching between flatband and dispersive bands \cite{PhysRevB.78.125104,PhysRevLett.121.263902}, such a system does not support any unbroken PT phase or completely real spectra once the non-Hermitian parameters are introduced, as generally encountered in non-Hermitian flatband systems \cite{PhysRevA.92.063813,PhysRevResearch.2.033127,Ge:18}. Meanwhile, peculiar features like secondary emissions unique to the non-Hermitian systems remain undiscovered in flatband lattices as far as we know. Thus, a natural question arises: can we establish a non-Hermitian flatband system that will exhibit (nonzero) threshold PT symmetry breaking, and how would the LZB oscillation dynamics change when the interplay between the flatband, EP and the driving field comes to play the role in such a system?

In this work, we propose a scheme to establish quasi-one-dimensional (1D) photonic PT-symmetric Lieb lattices (the Lieb ribbon), and to address the above questions by considering the LZB oscillations in the non-Hermitian flatband system. By introducing asymmetric complex-valued couplings and a refractive-index gradient perpendicular to the ribbon, we obtain the flatband lattice possessing the PT-symmetric phase and a triply degenerated higher-order EP. We show that, in such a non-Hermitian flatband system, PT symmetry breaking occurs only when the amplitude of gain/loss ratio exceeds a nonzero threshold. Moreover, an asymmetric energy distribution and pseudo-Hermitian propagation can take place for small non-Hermitian parameters. Importantly, we uncover that a wave-packet can become either amplified or damped during the LZB oscillations accompanied with strong secondary emissions in a non-Hermitian system.

A ribbon of Lieb photonic lattice composed of an array of periodically arranged evanescently coupled waveguides is shown in Fig.~\ref{Figure1}(a), which consists of five sites (\textbf{A}, \textbf{B}, \textbf{C}, \textbf{D} and \textbf{E}) per unit cell. Such a structure represents the quasi-1D form of two-dimensional Lieb lattices, which have attracted great interest in the past several years and have been employed to demonstrate fundamental flatband phenomena \cite{PhysRevLett.114.245503,PhysRevLett.114.245504,liu2020universal,Xia:16}. Although in a simple flatband lattice geometry, the Lieb ribbon allows to clarify the effects of the refractive index gradient fields on the flatband CLSs and LZB oscillations in a PT-symmetric non-Hermitian system. The external driven fields are applied by applying linear refractive index gradients either parallel ($x$-gradient) and perpendicular ($y$-gradient) to the ribbon. The non-Hermiticity is introduced by employing asymmetric complex-valued couplings ($t_{1}$ and $t_{2}$). Such non-Hermitian couplings can be realized by embedding gain or loss media between adjacent waveguides or periodically modulating the on-site gain/loss amplitude ratio \cite{PhysRevB.96.064305,PhysRevA.89.013848,PhysRevA.93.022102}. The Hamiltonian of the system in the tight-binding approximation can be written as:
\begin{figure}[htbp]
\centering
\includegraphics[width=9cm]{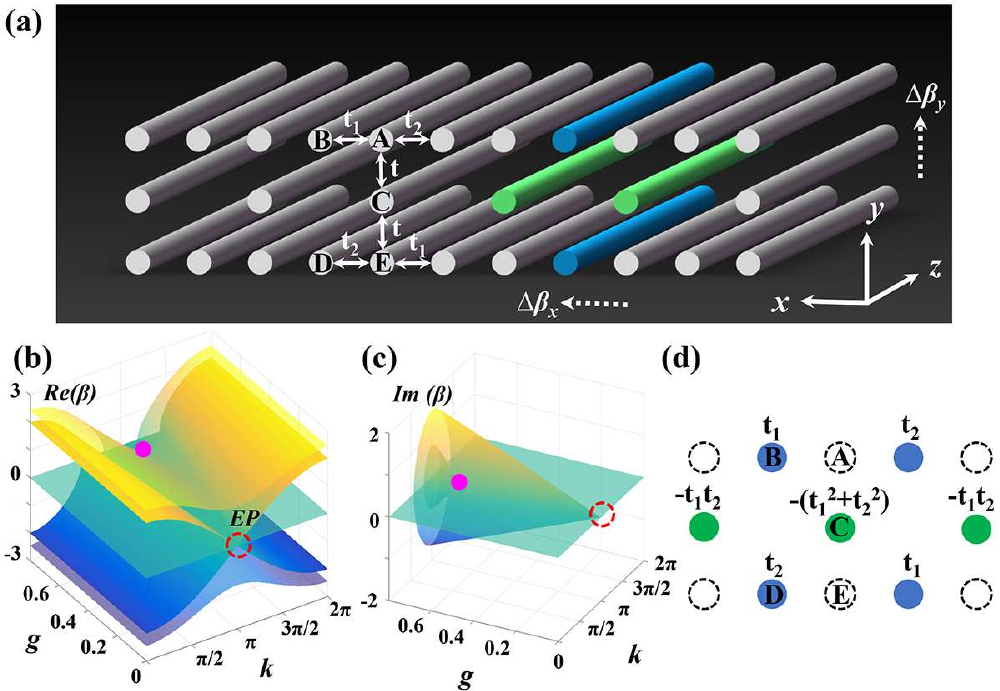}
\caption{(a) Schematic diagram of a driven Lieb lattice ribbon formed by coupled optical waveguide arrays with each unit cell consisting of five sites (\textbf{A}, \textbf{B}, \textbf{C}, \textbf{D} and \textbf{E}). The coupling coefficients are $t$, $t_{1}=t\left ( 1+ig\right )$ and $t_{2}=t\left ( 1-ig\right )$, where $g$ is the non-Hermitian parameter. Colored waveguides show the four-site CLS with equal amplitude but opposite phase of a uniform lattice (i.e., when $t_{1}=t_{2}$), ensuring destructive interference in the neighboring \textbf{A}(\textbf{E}) sites. $\Delta\beta_{x}$ and $\Delta\beta_{y}$ represent $x$- and $y$-gradient, respectively. (b, c) Calculated real and imaginary parts of the spectrum as a function of $g$ when no index gradient is present (i.e., $\Delta\beta_{x}=0$ , $\Delta\beta_{y}=0$). The red dashed circles represent the position of the EP, and the pink filled circles show the value in which the upmost and lowest dispersive bands become complex. (d) A flatband CLS occupies two plaquettes for $g\neq0$ with colored and filled circles representing the sites of nonzero energies, and the characters right next to them show the corresponding amplitudes. For simplicity, we set $t=1$, and the lattice period is 1.}.
\label{Figure1}
\end{figure}
\begin{equation}
\begin{aligned}
 H \!=\!&\sum\limits_n\!a_{n}^{\dagger }\left( t_{1}b_{n}+t_{2}b_{n-1}\!+\!tc_{n}\right)
     +e_{n}^{\dagger }\left(t_{2}d_{n}+\!t_{1}d_{n-1}+\!tc_{n}+\!h.c.\right)\\
     &+\Delta {\beta _y}\left( a_{n}^{\dagger }a_{n}+b_{n}^{\dagger }b_{n}-d_{n}^{\dagger }d_{n}-e_{n}^{\dagger }e_{n}\right)\\
     &+n\Delta {\beta _x}\left( a_{n}^{\dagger }a_{n}+b_{n}^{\dagger }b_{n}+c_{n}^{\dagger }c_{n}+d_{n}^{\dagger }d_{n}+e_{n}^{\dagger }e_{n}\right )\\
     &+\frac{\Delta \beta _{x}}{2}\left( b_{n}^{\dagger }b_{n}+d_{n}^{\dagger }d_{n}\right )
\end{aligned}
\end{equation}
where $a_{n}^{\dagger}$, $b_{n}^{\dagger}$, $c_{n}^{\dagger}$, $d_{n}^{\dagger}$, $e_{n}^{\dagger}$ and $a_{n}$, $b_{n}$, $c_{n}$, $d_{n}$, $e_{n}$ are the creation and annihilation operators in the n-th unit cell on the \textbf{A}, \textbf{B}, \textbf{C}, \textbf{D} and \textbf{E} sites, respectively. $t$, $t_{1}$ and $t_{2}$ are the coupling coefficients. We incorporate an arbitrary amount of PT-symmetric gain and loss into the system in a balanced way by setting $t_{1}=t\left ( 1+ig\right )$, $t_{2}=t\left ( 1-ig\right )$, where $g$ is the non-Hermitian parameter \cite{PhysRevB.96.064305,PhysRevResearch.2.033127}. Here $\Delta\beta_{x}$ and $\Delta\beta_{y}$ denote the wave-number difference between adjacent waveguides and thus define the $x$-gradient and $y$-gradient, respectively. The effective propagation constant (or the on-site energy) of the waveguides is determined by the refractive index gradient strength. We assume that the sites of the same unit cell have on-site energy difference $\Delta\beta_{y}$ along $y$ direction in the presence of $y$-gradient, while the effective propagation constant of the n-th unit cell is shifted by $\Delta\beta_{x}$ compared to the (n+1)-th unit cell for the $x$-gradient. Meanwhile, sublattices \textbf{A}, \textbf{C} and \textbf{E} of the n-th unit cell have the same effective propagation constant which is shifted by $\Delta\beta_{x}/2$ compared to the \textbf{B} (\textbf{D}) in the same unit cell.

This arrangement can be described by the discrete Schr{\"o}dinger equation $i\left ( d/dz\right )\Psi _{n}=H_{k}\Psi _{n}$, where $z$ denotes the propagation coordinate, $\Psi _{n}=\left ( a_{n}, b_{n}, c_{n} ,d_{n} ,e_{n}\right )^{T}$ is the five-component wave function describing the field amplitude in unit cell n, and the momentum space Hamiltonian $H_{k}$ is given by
\begin{equation}
H_{k}=\\
\begin{pmatrix}
  \Lambda _{A}&  t_{1}+t_{2}e^{-ik}& t & 0 & 0 \\
 t_{1}+t_{2}e^{ik}&  \Lambda _{B}& 0 & 0 &0 \\
 t&  0& \Lambda _{C} & 0 & t\\
 0& 0 & 0 &  \Lambda _{D} & t_{2}+t_{1}e^{ik} \\
 0& 0 & t &t_{2}+t_{1}e^{-ik}  &\Lambda _{E}
\end{pmatrix}
\end{equation}
($\Lambda _{A}=n\Delta\beta _{x}+\Delta \beta _{y}$, $\Lambda _{B}=\left ( n+1/2\right )\Delta\beta _{x}+\Delta \beta _{y}$, $\Lambda _{c}=n\Delta\beta _{x}$, $\Lambda _{D}=\left ( n+1/2\right )\Delta\beta _{x}-\Delta \beta _{y}$ and $\Lambda _{E}=n\Delta\beta _{x}-\Delta \beta _{y}$ are the gradients applied to the \textbf{A}, \textbf{B}, \textbf{C}, \textbf{D} and \textbf{E} sites, respectively.) Let us start from considering the case without the external field, i.e., $\Delta\beta_{y}=0$, $\Delta\beta_{y}=0$, and calculate the eigenvalues $\beta$ as a function of $g$. The real and imaginary parts are depicted in Figs.~\ref{Figure1}(b) and ~\ref{Figure1}(c), respectively. The Hermitian regime $g = 0$ (where both $t_{1}$ and $t_{2}$ are real) reveals that the lattice has five bands distributed symmetrically with respect to the central flatband $\beta=0$, which touches two dispersive bands at $k= \pi $  as marked by red circles. The system undergoes thresholdless PT symmetry breaking, i.e., the central two dispersive bands become complex for any value of $g\neq0$. Consequently, the band touching point is also the EP that representing the transition between the real and complex band regions. At the same time, band gap width of upmost and lowest dispersive bands decreases with the increasing of $g$. For $g=\sqrt{2}/2$, all the four dispersive bands are in the broken PT symmetry phase, as indicated by the pink filled circle. Nevertheless, it can be clearly found that the flatband $\beta = 0$ survives due to the chiral symmetry of the lattice \cite{PhysRevB.96.161104,PhysRevB.97.045120}. Moreover, different from uniform lattices ($t_{1}=t_{2}$) where a CLS occupies four sites in a single plaquette [Fig.~\ref{Figure1}(a)], the introduction of asymmetric couplings requires that a flatband CLS occupies at least two plaquettes. The irreducible CLS is sketched in Fig.~\ref{Figure1}(d) for $t=1$, with \textbf{B}(\textbf{D}) and \textbf{C} sites have unequal amplitudes and opposite phase, ensuring destructive interference in the neighboring \textbf{A}(\textbf{E}) sites. Note that such a state becomes overlapping of two fundamental CLSs in Fig.~\ref{Figure1}(a) for $g=0$ ($t_{1}=t_{2}=t=1$).
\begin{figure}[htbp]
\centering
\includegraphics[width=9cm]{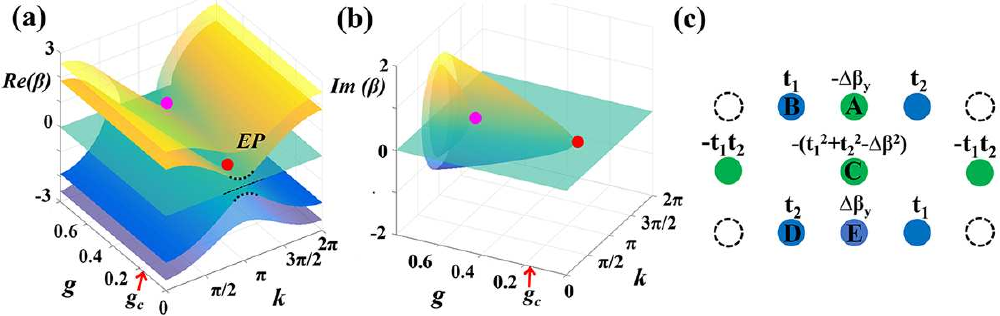}
\caption{Same as in Figs.~\ref{Figure1}(b-d) but with a $y$-gradient $\Delta\beta_{y}=0.3t$ ($t = 1$). Due to the presence of $y$-gradient field, two symmetric gaps open for $g=0$ and a triply degenerated EP (marked by red filled circles) with nonzero value appears as $g$ is increased to $g_{c}=0.17$. The new CLS occupies two plaquettes, where extra \textbf{A} and \textbf{E} sites are included, with opposite phases (represented by blue and green colors) and same amplitude equalling field strength $\Delta\beta_{y}$.}
\label{Figure2}
\end{figure}
Next, we consider the case where a $y$-gradient is applied. Figures.~\ref{Figure2}(a) and ~\ref{Figure2}(b) show the typical spectrum for $\Delta\beta_{y}=0.3t$ ($t = 1$, and thus $\Delta\beta_{y}=0.3$). The band degeneracy is lifted with two symmetric gaps emerge between the flatband and dispersive bands for $g = 0$. With the increase of the non-Hermitian parameter $g$, the band gap decreases and coincides with the PT symmetry breaking transition for critical value $g_{c}=0.5\sqrt{1-\Delta \beta _{y}^{2}-\sqrt{1-4\Delta \beta _{y}^{2}}}=0.17$, at which the eigenmodes of the dispersive bands coalesce into a degenerated one. More importantly, the flatband still survives, and different from previously studied non-Hermitian systems, here the EP becomes triply degenerated in virtue of the existence of flatband. Accordingly, all the eigenvalues are completely real for $g<g_{c}$, and a flatband lattice with nonzero threshold PT symmetry breaking is obtained. For $g>g_{c}$ the system turns into the symmetry-broken phase. As $g$ is further  increased to cross the value $g=0.5\sqrt{1-\Delta \beta _{y}^{2}+\sqrt{1-4\Delta \beta _{y}^{2}}}=0.65$, the highest and lowest bands merge and two pairs of complex-conjugates eigenvalues are observed. Furthermore, the CLS shown in Fig.~\ref{Figure1}(a) is no longer an exact solution of the lattice, as for $\Delta\beta_{y}\neq0$ non-zero amplitudes appear in sites \textbf{A} and \textbf{E} as well as the $y$-gradient remodulates the amplitude in the central \textbf{C} site  [Fig.~\ref{Figure2}(c)].

\begin{figure}[H]
\centering
\includegraphics[width=9cm]{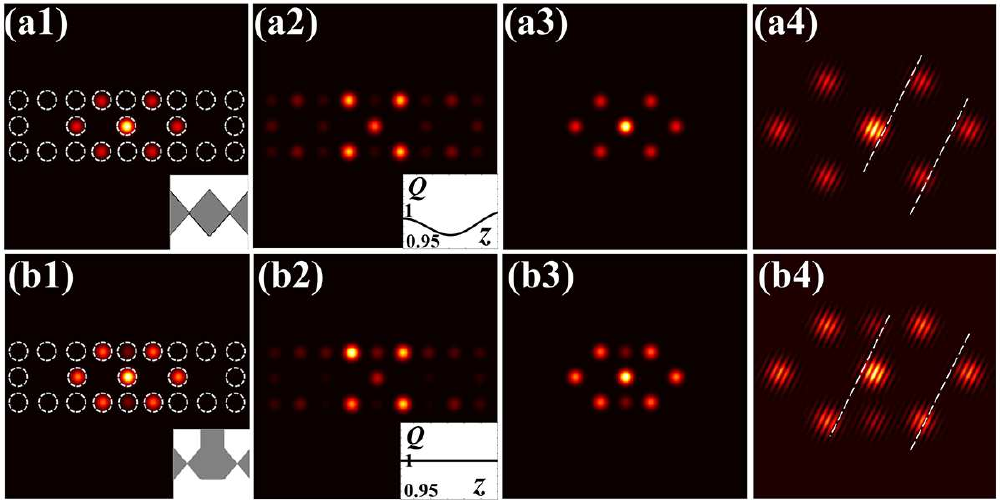}
\caption{Evolution dynamics of the flatband CLSs in PT-symmetric photonic lattices (first row) without refractive index gradients $\Delta\beta_{x}=0$, $\Delta\beta_{y}=0$ and (second row) with a $y$-gradient  $\Delta\beta_{y}=0.5t$. First column: input beams. Insets show the phase structure, where white region represents $\pi$ phase and gray area represents 0 phase. Second column: in-phase output with discrete diffraction appears. Insets show the evolution of quasi-energy Q, which is conserved in (b2) as the system is operated below the PT symmetry threshold value. Third column: out-of-phase output with each image normalized so that the total intensity is 1. Though the system is non-Hermitian, the original excitations do not lead to any diffraction and the total energy remains constant. Fourth column: corresponding  zoom-in interferograms of the third column. We set $t$=250 mm$^{-1}$, $g=0.3$, and propagation length $z$=6 mm. }
\label{Figure3}
\end{figure}

To verify our analytical results, we numerically simulate the evolution dynamics of CLSs for  $g=0.3$ with propagation distance $z$=6 mm. We consider 260 unit cells of waveguide arrays arranged according to Fig.~\ref{Figure1} and excite the central two plaquettes. Figures.~\ref{Figure3}(a1) and ~\ref{Figure3}(b1) show the intensity and phase of the input beam $\Psi _{n}$ similar to that of flatband CLSs shown in Figs.~\ref{Figure1}(d) and ~\ref{Figure2}(c), respectively. The first row of Fig.~\ref{Figure3} shows the corresponding results of lattices without the refractive index gradient. It can be clearly seen that the initial excitation propagates without any diffraction due to the excitation of flatband modes [Fig.~\ref{Figure3}(a3)]. Furthermore, the initial intensity remains constant and consequently is conserved even though the system is in a PT symmetry broken phase. Moreover, the phase measurement obtained by interfering the output with an inclined plane wave further confirms that the initial phase structure is well preserved [Fig.~\ref{Figure3}(a4)]. Instead, if the input beam has equal phase, a discrete diffraction pattern is obtained [Fig.~\ref{Figure3}(a2)] and the power is not conserved as shown in the inset. Similar results can be found for the case $\Delta\beta_{y}\neq0$. For better visibility, we set $\Delta\beta_{y}=0.5t$ (\textbf{A} and \textbf{E} sites become more visible). The excitation [Fig.~\ref{Figure3}(b1)] now contains two more sites and stays invariant  during propagation [Fig.~\ref{Figure3}(b3)]. However, due to the presence of $y$-gradient, the discrete diffraction of equal phase input becomes asymmetrical [Fig.~\ref{Figure3}(b2)]. It should be noted that the system is in PT-symmetry unbroken phase as the critical value for the PT transition is set to $g_{c}=0.43$ in this case. Consequently, the quasi-energy $Q=\sum_{n}\Psi \left ( n\right )\Psi ^{\ast }\left ( -n\right )$ is also conserved in Fig.~\ref{Figure3}(b2) \cite{PhysRevLett.103.123601}.

\begin{figure}[H]
\centering
\includegraphics[width=9cm]{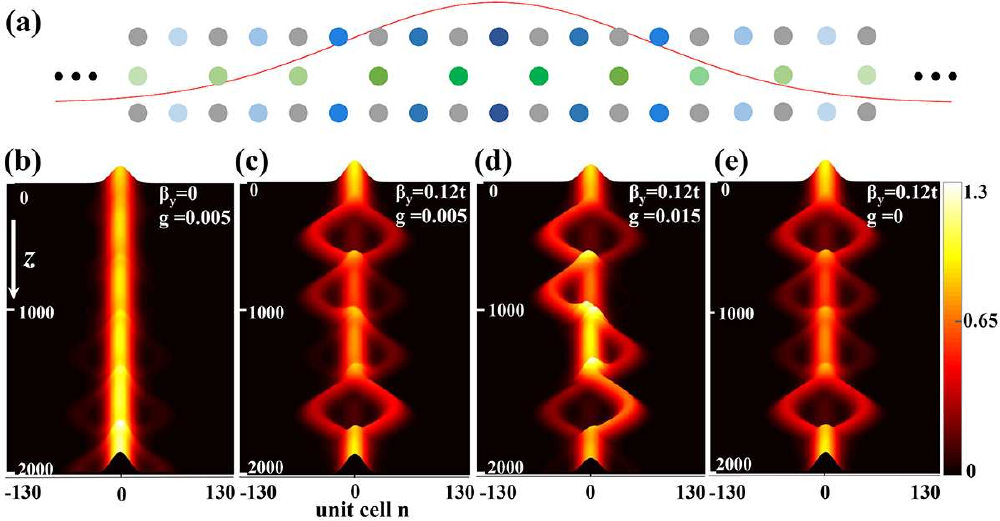}
\caption{Flatband LZB oscillations with an $x$-gradient $\Delta\beta_{x}=0.05t$. (a) An initial excitation beam formed by combining the CLSs with a broad Gaussian distribution (red line)  $b_{n}=-c_{n}=d_{n}=e^{-n^{2}/2\sigma ^{2}}$ ($\sigma=15$) at normal incidence. Colored sites represent nonzero amplitudes with blue and green sites having $\pi$ phase difference. (b) $\Delta\beta_{y}=0$, $g=0.005$, the input keeps compactness with slight and symmetric oscillations for non-Hermitian lattices in the absence of $y$-gradient. (c) $\Delta\beta_{y}=0.12t$, $g=0.005$, the total energy stays constant over many periods and the energy distributes asymmetrically for non-Hermitian lattices in the presence of $y$-gradient.  (d) $\Delta\beta_{y}=0.12t$, $g=0.015$, the energy of the wave-packet is amplified during oscillations. (e) $\Delta\beta_{y}=0.12t$, $g=0$, initial state evolves in a symmetric way for Hermitian lattices in the presence of $y$-gradient for direct comparison. The oscillation period is $l=2\pi /\Delta \beta_{x} $ and the propagation length $L=5l$ is divided into 2000 steps. For all figures, the input peak intensity is normalized to 1.}
\label{Figure4}
\end{figure}

Now, we consider the case for $\Delta\beta_{x}\neq0$ and investigate the flatband CLSs dynamics in the PT-symmetric Lieb lattice ribbon. In our platform, applying a constant refractive index gradient parallel to the lattice is similar to a constant dc electric field applied to a crystal, and thus resulting in Bloch oscillations of the wave-packet. Linear combinations of CLSs shown in Fig.~\ref{Figure1}(a) with a broad Gaussian distribution $b_{n}=-c_{n}=d_{n}=e^{-n^{2}/2\sigma ^{2}}$ ($\sigma=15$) is set as an input excitation at normal incidence [Fig.~\ref{Figure4}(a)]. Different from the rhombic lattices where flatband and CLSs do not change in the presence of $x$-gradient \cite{PhysRevLett.116.245301,XiaAPL}, it is easy to find that the input CLSs no longer stay strictly localized for $\Delta\beta_{x}\neq0$. However, the central band remains nearly flat when $x$-gradient is small, and the input still can be used to excite flatband \cite{long2017topological}. Figures.~\ref{Figure4}(b-e) depict the numerical results with $\Delta\beta_{x}=0.05t$.
\noindent For the non-Hermitian case $\Delta\beta_{y}=0$, $g=0.005$, the input remains almost invariant with only sightly symmetric Bloch oscillations observed [Fig.~\ref{Figure4}(b)]. In this case, the spectrum is symmetric and the gap width is zero [see Fig.~\ref{Figure1}(b)], and the initial state ceases to be a linear combination of CLSs depicted in Fig.~\ref{Figure1}(d) with a part of the wave-packet splitting into dispersive bands. In addition, the system is PT symmetry broken and therefore the energy is obviously amplified. Noticeably different features can be found for the cases of $\Delta\beta_{y}\neq0$, $g\neq0$ [Figs.~\ref{Figure4}(c) and ~\ref{Figure4}(d)]. More specifically, we observe sudden switches from an almost unaltered state into an oscillating pattern and Landau-Zener tunneling due to the triply interband transitions. The beam intensity starts to exhibit an asymmetric energy distribution with the increase of g. This asymmetry is likely attributed to the asymmetry of the energy spectrum, i.e., the spectrum become asymmetric with the simultaneous appearance of $x$-gradient, $y$-gradient and non-Hermitian coupling, which is similar to the cases with magnetic field in the Hermitian systems \cite{PhysRevLett.116.245301}. Note that, compared with the case in Fig.~\ref{Figure2}, the Wannier-Stark ladder spectrum is complex-valued even though $g< g_{c}$ ($\Delta\beta_{y}=0.12t$, and the critical value at EP is $g_{c}= 0.06$) \cite{PhysRevLett.103.123601,PhysRevLett.124.066602,wimmer2015obser,xu2016experimental}. This is because the external $x$-gradient field breaks the PT symmetry of the Hamiltonian. Remarkably, we find that a pseudo-Hermitian propagation is restored for very small values of g ($g\ll g_{c}$), where the average total power remains constant over many periods of Bloch oscillations, see Fig.~\ref{Figure4}(c) for $g=0.005$. Meanwhile, due to the complex band structure, the wave-packet is amplified each time passing through the EP for bigger value of g, see Fig.~\ref{Figure4}(d) for $g=0.015$. For comparison, we also investigate the Hermitian case in the presence of $y$-gradient in Fig.~\ref{Figure4}(e) ($\Delta\beta_{y}\neq0$, $g=0$). In this case, the initial state is conserved and it evolves in a symmetric way, reflecting the band symmetry shown in Fig.~\ref{Figure2}(a) for $g=0$.
\begin{figure}[H]
\centering
\includegraphics[width=9cm]{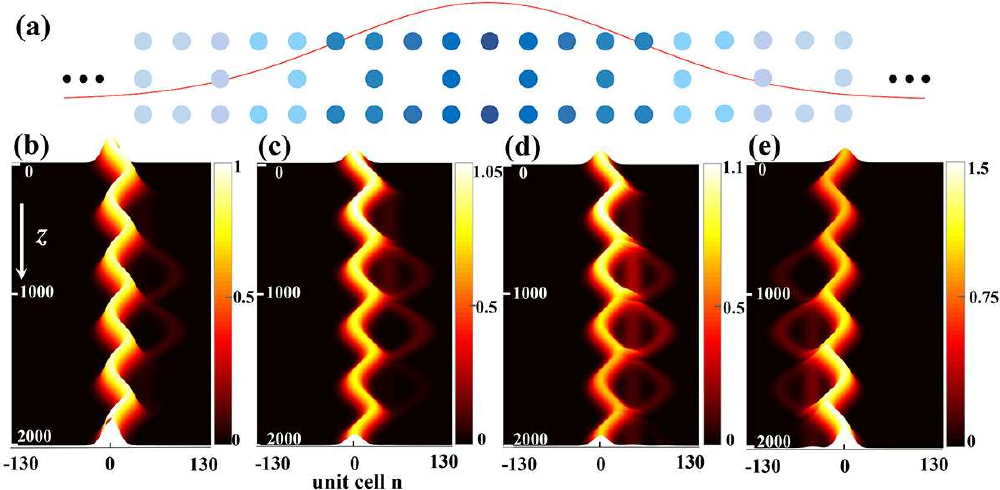}
\caption{LZB oscillations under a dispersive band excitation. (a) An initial input beam formed by combining the CLSs with a broad Gaussian distribution (red line) $a_{n}=b_{n}=c_{n}=d_{n}=e_{n}=e^{-n^{2}/2\sigma ^{2}}$ ($\sigma=15$) at normal incidence. Colored sites represent nonzero amplitudes with equal phase structure. (b) Oscillations in a Hermitian system with $g=0$, showing constant amplitude during propagation.  (c, d) Oscillations in a non-Hermitian system with (c) $g=0.015$ and (d) $g=0.03$, showing secondary emissions and a decaying amplitude.  (e) same as in (d) but the sign of $x$-gradient field is opposite, showing a growing amplitude. Other parameters are the same as in Fig.~\ref{Figure4}(c). }
\label{Figure5}
\end{figure}
One of the most striking characteristics of non-Hermitian Bloch oscillations is the existence of the so-called secondary emission that occurs at the EP \cite{wimmer2015obser,xu2016experimental}.  The degeneracy at the EP causes a redistribution of the initial excitation. Therefore, each time the wave-packet passes through the EP, a new branch of Bloch oscillations appears. However, such a phenomenon is not obvious when the central flatband is excited. To reveal this effect, we simulate the Bloch oscillations of the dispersive bands. Different form the excitation in Fig.~\ref{Figure4}, the input beam has equal phase structure and all the sublattices including \textbf{A} and \textbf{E} are excited ($a_{n}=b_{n}=c_{n}=d_{n}=e_{n}=e^{-n^{2}/2\sigma ^{2}}$) [Fig.~\ref{Figure5}(a)] . In the Hermitian case $g=0$ [Fig.~\ref{Figure5}(b)], this initial excitation yields sinusoidal energy oscillations together with only weak Landau-Zener tunneling between two dispersive bands. Moreover, no energy distributes in flatband. The situation nevertheless changes drastically in non-Hermitian regime, as for $g=0.015$ the energy starts to evolve into the flatband [Fig.~\ref{Figure5}(c)]. Such an effect appears more clearly with the increase of g, as the secondary emissions occur each time when the oscillatory motion passes the triply degenerated EP, leading to distinct energy distribution into the flatband and strong energy tunneling between the flatband and dispersive bands [Fig.~\ref{Figure5}(d)]. Let us notice that in both cases Figs.~\ref{Figure5}(c) and ~\ref{Figure5}(d) the total energy decreases along the propagation. This can be changed so that the total energy becomes amplified along the propagation while preserving the oscillations observed in Fig.~\ref{Figure5}(d) simply by reversing the sign of the $x$-gradient field - as illustrated in Fig.~\ref{Figure5}(e). These results indicate that the wave-packet can become either amplified or damped during the LZB oscillations accompanied with strong secondary emissions, depending on the direction of the driving field (i.e., the applied index gradient as in our photonic lattices) in the non-Hermitian system \cite{PhysRevLett.103.123601}.

We have theoretically and numerically investigated the LZB oscillations in flatband photonic Lieb ribbons with nonzero threshold PT symmetry breaking. The PT symmetry is introduced by judiciously tailored couplings with asymmetric complex coupling coefficients, and the refractive-index gradient is applied acting as an external field. In such a system, the flatband can survive and a nonzero triply degenerated EP can be established in the presence of $y$-gradient. Furthermore, the LZB oscillations exhibit intriguing features such as asymmetric energy distribution, pseudo-Hermitian propagation, and secondary emissions at the EP. The proposed artificial lattices can also be realized and other physical platforms, where the presence of flatband in a PT-symmetric phase could lead to intriguing fundamental phenomena when disorder and nonlinearity are introduced. In future works, it could also be interesting to investigate such higher-order EPs in two-dimensional systems and non-Hermitian topological transitions associated with them.


\acknowledgments
{\it Acknowledgments.---}
This work was supported by the National Key R$\&$D Program of China (Grant No. 2017YFA0303800), the NSFC (Grant Nos. 11922408, 11704102, and 91750204), the 111 Project (Grant No. B07013).

\bibliographystyle{unsrt}
\bibliography{reff}

\begin{thebibliography}{10}

\bibitem{PhysRevLett.106.236802}
Evelyn Tang, Jia-Wei Mei, and Xiao-Gang Wen.
\newblock High-temperature fractional quantum {H}all states.
\newblock {\em Phys. Rev. Lett.}, 106:236802, Jun 2011.

\bibitem{PhysRevLett.106.236803}
Kai Sun, Zhengcheng Gu, Hosho Katsura, and S.~Das~Sarma.
\newblock Nearly flatbands with nontrivial topology.
\newblock {\em Phys. Rev. Lett.}, 106:236803, Jun 2011.

\bibitem{PhysRevLett.106.236804}
Titus Neupert, Luiz Santos, Claudio Chamon, and Christopher Mudry.
\newblock Fractional quantum {H}all states at zero magnetic field.
\newblock {\em Phys. Rev. Lett.}, 106:236804, Jun 2011.

\bibitem{PhysRevB.83.220503}
N.~B. Kopnin, T.~T. Heikkil\"a, and G.~E. Volovik.
\newblock High-temperature surface superconductivity in topological flat-band
  systems.
\newblock {\em Phys. Rev. B}, 83:220503, Jun 2011.

\bibitem{PhysRevA.87.061803}
Rodrigo~A. Vicencio and Magnus Johansson.
\newblock Discrete flat-band solitons in the kagome lattice.
\newblock {\em Phys. Rev. A}, 87:061803, Jun 2013.

\bibitem{PhysRevE.92.032912}
Magnus Johansson, Uta Naether, and Rodrigo~A. Vicencio.
\newblock Compactification tuning for nonlinear localized modes in sawtooth
  lattices.
\newblock {\em Phys. Rev. E}, 92:032912, Sep 2015.

\bibitem{PhysRevA.96.063838}
P.~P. Beli\ifmmode~\check{c}\else \v{c}\fi{}ev,
  G.~Gligori\ifmmode~\acute{c}\else \'{c}\fi{}, A.~Maluckov,
  M.~Stepi\ifmmode~\acute{c}\else \'{c}\fi{}, and M.~Johansson.
\newblock Localized gap modes in nonlinear dimerized {L}ieb lattices.
\newblock {\em Phys. Rev. A}, 96:063838, Dec 2017.

\bibitem{DanieliLTP}
C.~Danieli, A.~Maluckov, and S.~Flach.
\newblock Compact discrete breathers on flat-band networks.
\newblock {\em Low Temperature Physics}, 44(7):678--687, 2018.

\bibitem{cao2018correlated}
Yuan Cao, Valla Fatemi, Ahmet Demir, Shiang Fang, Spencer~L Tomarken, Jason~Y
  Luo, Javier~D Sanchez-Yamagishi, Kenji Watanabe, Takashi Taniguchi, Efthimios
  Kaxiras, Ray~C Ashoori, and Pablo Jarillo-Herrero.
\newblock Correlated insulator behaviour at half-filling in magic-angle
  graphene superlattices.
\newblock {\em Nature}, 556(7699):80--84, 2018.

\bibitem{balents2020supercond}
Leon Balents, Cory~R Dean, Dmitri~K Efetov, and Andrea~F Young.
\newblock Superconductivity and strong correlations in moir{\'e} flat bands.
\newblock {\em Nat. Phys.}, 16(7):725--733, 2020.

\bibitem{wang2020localization}
Peng Wang, Yuanlin Zheng, Xianfeng Chen, Changming Huang, Yaroslav~V Kartashov,
  Lluis Torner, Vladimir~V Konotop, and Fangwei Ye.
\newblock Localization and delocalization of light in photonic moir{\'e}
  lattices.
\newblock {\em Nature}, 577(7788):42--46, 2020.

\bibitem{PhysRevLett.114.245503}
Rodrigo~A. Vicencio, Camilo Cantillano, Luis Morales-Inostroza, Basti\'an Real,
  Cristian Mej\'{\i}a-Cort\'es, Steffen Weimann, Alexander Szameit, and
  Mario~I. Molina.
\newblock Observation of localized states in {L}ieb photonic lattices.
\newblock {\em Phys. Rev. Lett.}, 114:245503, Jun 2015.

\bibitem{PhysRevLett.114.245504}
Sebabrata Mukherjee, Alexander Spracklen, Debaditya Choudhury, Nathan Goldman,
  Patrik \"Ohberg, Erika Andersson, and Robert~R. Thomson.
\newblock Observation of a localized flat-band state in a photonic {L}ieb
  lattice.
\newblock {\em Phys. Rev. Lett.}, 114:245504, Jun 2015.

\bibitem{Xia:16}
Shiqiang Xia, Yi~Hu, Daohong Song, Yuanyuan Zong, Liqin Tang, and Zhigang Chen.
\newblock Demonstration of flat-band image transmission in optically induced
  {L}ieb photonic lattices.
\newblock {\em Opt. Lett.}, 41(7):1435--1438, Apr 2016.

\bibitem{Zong16}
Yuanyuan Zong, Shiqiang Xia, Liqin Tang, Daohong Song, Yi~Hu, Yumiao Pei, Jing
  Su, Yigang Li, and Zhigang Chen.
\newblock Observation of localized flat-band states in {K}agome photonic
  lattices.
\newblock {\em Opt. Express}, 24(8):8877--8885, Apr 2016.

\bibitem{Derzhko2015}
Oleg Derzhko, Johannes Richter, and Mykola Maksymenko.
\newblock Strongly correlated flat-band systems: The route from {H}eisenberg
  spins to {H}ubbard electrons.
\newblock {\em International Journal of Modern Physics B}, 29(12):1530007,
  2015.

\bibitem{Leykam2018}
Daniel Leykam, Alexei Andreanov, and Sergej Flach.
\newblock Artificial flat band systems: from lattice models to experiments.
\newblock {\em Advances in Physics: X}, 3(1):1473052, 2018.

\bibitem{LeykamAPL}
Daniel Leykam and Sergej Flach.
\newblock Perspective: Photonic flatbands.
\newblock {\em APL Photonics}, 3(7):070901, 2018.

\bibitem{tang2020photonic}
Liqin Tang, Daohong Song, Shiqi Xia, Shiqiang Xia, Jina Ma, Wenchao Yan, Yi~Hu,
  Jingjun Xu, Daniel Leykam, and Zhigang Chen.
\newblock Photonic flat-band lattices and unconventional light localization.
\newblock {\em Nanophotonics}, 9(5):1161--1176, 2020.

\bibitem{Vicencio}
Rodrigo A.~Vicencio Poblete.
\newblock Photonic flat band dynamics.
\newblock {\em Advances in Physics: X}, 6(1):1878057, 2021.

\bibitem{PhysRevLett.121.075502}
Sebabrata Mukherjee, Marco Di~Liberto, Patrik \"Ohberg, Robert~R. Thomson, and
  Nathan Goldman.
\newblock Experimental observation of {A}haronov-{B}ohm cages in photonic
  lattices.
\newblock {\em Phys. Rev. Lett.}, 121:075502, Aug 2018.

\bibitem{kremer2020square}
Mark Kremer, Ioannis Petrides, Eric Meyer, Matthias Heinrich, Oded Zilberberg,
  and Alexander Szameit.
\newblock A square-root topological insulator with non-quantized indices
  realized with photonic {A}haronov-{B}ohm cages.
\newblock {\em Nat. Commun.}, 11(1):1--6, 2020.

\bibitem{feng2017non}
Liang Feng, Ramy El-Ganainy, and Li~Ge.
\newblock Non-{H}ermitian photonics based on parity--time symmetry.
\newblock {\em Nat. Photonics}, 11(12):752--762, 2017.

\bibitem{el2018non}
Ramy El-Ganainy, Konstantinos~G Makris, Mercedeh Khajavikhan, Ziad~H
  Musslimani, Stefan Rotter, and Demetrios~N Christodoulides.
\newblock Non-{H}ermitian physics and {PT} symmetry.
\newblock {\em Nat. Phys.}, 14(1):11--19, 2018.

\bibitem{ozdemir2019parity}
{\c{S}}ahin~Kaya {\"O}zdemir, Stefan Rotter, Franco Nori, and L~Yang.
\newblock Parity--time symmetry and exceptional points in photonics.
\newblock {\em Nat. Mater.}, 18(8):783--798, 2019.

\bibitem{PhysRevLett.106.213901}
Zin Lin, Hamidreza Ramezani, Toni Eichelkraut, Tsampikos Kottos, Hui Cao, and
  Demetrios~N. Christodoulides.
\newblock Unidirectional invisibility induced by
  $\mathcal{P}\mathcal{T}$-symmetric periodic structures.
\newblock {\em Phys. Rev. Lett.}, 106:213901, May 2011.

\bibitem{PhysRevLett.115.040402}
Julia~M. Zeuner, Mikael~C. Rechtsman, Yonatan Plotnik, Yaakov Lumer, Stefan
  Nolte, Mark~S. Rudner, Mordechai Segev, and Alexander Szameit.
\newblock Observation of a topological transition in the bulk of a
  non-{H}ermitian system.
\newblock {\em Phys. Rev. Lett.}, 115:040402, Jul 2015.

\bibitem{xia2021nonlinear}
Shiqi Xia, Dimitrios Kaltsas, Daohong Song, Ioannis Komis, Jingjun Xu,
  Alexander Szameit, Hrvoje Buljan, Konstantinos~G Makris, and Zhigang Chen.
\newblock Nonlinear tuning of {PT} symmetry and non-{H}ermitian topological
  states.
\newblock {\em Science}, 372(6537):72--76, 2021.

\bibitem{Feng2014single}
Liang Feng, Zi~Jing Wong, Ren-Min Ma, Yuan Wang, and Xiang Zhang.
\newblock Single-mode laser by parity-time symmetry breaking.
\newblock {\em Science}, 346(6212):972--975, 2014.

\bibitem{hodaei2014parity}
Hossein Hodaei, Mohammad-Ali Miri, Matthias Heinrich, Demetrios~N
  Christodoulides, and Mercedeh Khajavikhan.
\newblock Parity-time--symmetric microring lasers.
\newblock {\em Science}, 346(6212):975--978, 2014.

\bibitem{hodaei2017}
Hossein Hodaei, Absar~U Hassan, Steffen Wittek, Hipolito Garcia-Gracia, Ramy
  El-Ganainy, Demetrios~N Christodoulides, and Mercedeh Khajavikhan.
\newblock Enhanced sensitivity at higher-order exceptional points.
\newblock {\em Nature}, 548(7666):187--191, 2017.

\bibitem{delplace2021}
Pierre Delplace, Tsuneya Yoshida, and Yasuhiro Hatsugai.
\newblock Symmetry-protected higher-order exceptional points and their
  topological characterization, arXiv:2103.08232, 2021.

\bibitem{mandal2021}
Ipsita Mandal and Emil~J. Bergholtz.
\newblock Symmetry and higher-order exceptional points, arXiv:2103.15729, 2021.

\bibitem{PhysRevB.96.064305}
Daniel Leykam, Sergej Flach, and Y.~D. Chong.
\newblock Flat bands in lattices with non-{H}ermitian coupling.
\newblock {\em Phys. Rev. B}, 96:064305, Aug 2017.

\bibitem{PhysRevLett.120.093901}
Bingkun Qi, Lingxuan Zhang, and Li~Ge.
\newblock Defect states emerging from a non-{H}ermitian flatband of photonic
  zero modes.
\newblock {\em Phys. Rev. Lett.}, 120:093901, Feb 2018.

\bibitem{PhysRevA.96.011802}
Hamidreza Ramezani.
\newblock Non-{H}ermiticity-induced flat band.
\newblock {\em Phys. Rev. A}, 96:011802, Jul 2017.

\bibitem{PhysRevA.99.033810}
L.~Jin.
\newblock Flat band induced by the interplay of synthetic magnetic flux and
  non-{H}ermiticity.
\newblock {\em Phys. Rev. A}, 99:033810, Mar 2019.

\bibitem{PhysRevLett.123.183601}
Tobias Biesenthal, Mark Kremer, Matthias Heinrich, and Alexander Szameit.
\newblock Experimental realization of $\mathcal{P}\mathcal{T}$-symmetric flat
  bands.
\newblock {\em Phys. Rev. Lett.}, 123:183601, Oct 2019.

\bibitem{bloch1928quantum}
Felix Bloch.
\newblock Quantum mechanics of electrons in crystal lattices.
\newblock {\em Z. Phys}, 52:555--600, 1928.

\bibitem{zener1934theory}
Clarence Zener.
\newblock A theory of the electrical breakdown of solid dielectrics.
\newblock {\em Proc. R. Soc. Lond. A}, 145(855):523--529, 1934.

\bibitem{PhysRevLett.83.4752}
T.~Pertsch, P.~Dannberg, W.~Elflein, A.~Br\"auer, and F.~Lederer.
\newblock Optical {B}loch oscillations in temperature tuned waveguide arrays.
\newblock {\em Phys. Rev. Lett.}, 83:4752--4755, Dec 1999.

\bibitem{PhysRevLett.83.4756}
R.~Morandotti, U.~Peschel, J.~S. Aitchison, H.~S. Eisenberg, and Y.~Silberberg.
\newblock Experimental observation of linear and nonlinear optical {B}loch
  oscillations.
\newblock {\em Phys. Rev. Lett.}, 83:4756--4759, Dec 1999.

\bibitem{PhysRevLett.91.263902}
Riccardo Sapienza, Paola Costantino, Diederik Wiersma, Mher Ghulinyan,
  Claudio~J. Oton, and Lorenzo Pavesi.
\newblock Optical analogue of electronic {B}loch oscillations.
\newblock {\em Phys. Rev. Lett.}, 91:263902, Dec 2003.

\bibitem{corrielli2013fractional}
Giacomo Corrielli, Andrea Crespi, Giuseppe Della~Valle, Stefano Longhi, and
  Roberto Osellame.
\newblock Fractional {B}loch oscillations in photonic lattices.
\newblock {\em Nature communications}, 4(1):1--6, 2013.

\bibitem{PhysRevLett.96.053903}
Henrike Trompeter, Wieslaw Krolikowski, Dragomir~N. Neshev, Anton~S.
  Desyatnikov, Andrey~A. Sukhorukov, Yuri~S. Kivshar, Thomas Pertsch, Ulf
  Peschel, and Falk Lederer.
\newblock Bloch oscillations and {Z}ener tunneling in two-dimensional photonic
  lattices.
\newblock {\em Phys. Rev. Lett.}, 96:053903, Feb 2006.

\bibitem{PhysRevLett.121.033904}
Yong Sun, Daniel Leykam, Stephen Nenni, Daohong Song, Hong Chen, Y.~D. Chong,
  and Zhigang Chen.
\newblock Observation of valley {L}andau-{Z}ener-{B}loch oscillations and
  pseudospin imbalance in photonic graphene.
\newblock {\em Phys. Rev. Lett.}, 121:033904, Jul 2018.

\bibitem{PhysRevLett.116.245301}
Ramaz Khomeriki and Sergej Flach.
\newblock Landau-{Z}ener {B}loch oscillations with perturbed flat bands.
\newblock {\em Phys. Rev. Lett.}, 116:245301, Jun 2016.

\bibitem{long2017topological}
Yang Long and Jie Ren.
\newblock Topological {L}andau-{Z}ener {B}loch oscillations in photonic
  {F}loquet {L}ieb lattices.
\newblock {\em arXiv:1706.01107}, 2017.

\bibitem{XiaAPL}
Shiqiang Xia, Carlo Danieli, Wenchao Yan, Denghui Li, Shiqi Xia, Jina Ma, Hai
  Lu, Daohong Song, Liqin Tang, Sergej Flach, and Zhigang Chen.
\newblock Observation of quincunx-shaped and dipole-like flatband states in
  photonic rhombic lattices without band-touching.
\newblock {\em APL Photon.}, 5(1):016107, 2020.

\bibitem{PhysRevLett.103.123601}
S.~Longhi.
\newblock Bloch oscillations in complex crystals with $\mathcal{P}\mathcal{T}$
  symmetry.
\newblock {\em Phys. Rev. Lett.}, 103:123601, Sep 2009.

\bibitem{PhysRevLett.124.066602}
S.~Longhi.
\newblock Non-{B}loch-band collapse and chiral {Z}ener tunneling.
\newblock {\em Phys. Rev. Lett.}, 124:066602, Feb 2020.

\bibitem{wimmer2015obser}
Martin Wimmer, Mohammed-Ali Miri, Demetrios Christodoulides, and Ulf Peschel.
\newblock Observation of {B}loch oscillations in complex {PT}-symmetric
  photonic lattices.
\newblock {\em Sci. Rep.}, 5(1):1--8, 2015.

\bibitem{xu2016experimental}
Ye-Long Xu, William~S Fegadolli, Lin Gan, Ming-Hui Lu, Xiao-Ping Liu, Zhi-Yuan
  Li, Axel Scherer, and Yan-Feng Chen.
\newblock Experimental realization of {B}loch oscillations in a parity-time
  synthetic silicon photonic lattice.
\newblock {\em Nat. Commun.}, 7(1):1--6, 2016.

\bibitem{PhysRevA.103.023721}
J.~Ramya~Parkavi, V.~K. Chandrasekar, and M.~Lakshmanan.
\newblock Stable {B}loch oscillations and {L}andau-{Z}ener tunneling in a
  non-{H}ermitian $\mathcal{PT}$-symmetric flat-band lattice.
\newblock {\em Phys. Rev. A}, 103:023721, Feb 2021.

\bibitem{PhysRevB.78.125104}
Doron~L. Bergman, Congjun Wu, and Leon Balents.
\newblock Band touching from real-space topology in frustrated hopping models.
\newblock {\em Phys. Rev. B}, 78:125104, Sep 2008.

\bibitem{PhysRevLett.121.263902}
Shiqi Xia, Ajith Ramachandran, Shiqiang Xia, Denghui Li, Xiuying Liu, Liqin
  Tang, Yi~Hu, Daohong Song, Jingjun Xu, Daniel Leykam, Sergej Flach, and
  Zhigang Chen.
\newblock Unconventional flatband line states in photonic {L}ieb lattices.
\newblock {\em Phys. Rev. Lett.}, 121:263902, Dec 2018.

\bibitem{PhysRevA.92.063813}
Mario~I. Molina.
\newblock Flat bands and $\mathcal{PT}$ symmetry in quasi-one-dimensional
  lattices.
\newblock {\em Phys. Rev. A}, 92:063813, Dec 2015.

\bibitem{PhysRevResearch.2.033127}
S.~M. Zhang and L.~Jin.
\newblock Localization in non-{H}ermitian asymmetric rhombic lattice.
\newblock {\em Phys. Rev. Research}, 2:033127, Jul 2020.

\bibitem{Ge:18}
Li~Ge.
\newblock Non-{H}ermitian lattices with a flat band and polynomial power
  increase.
\newblock {\em Photon. Res.}, 6(4):A10, Apr 2018.

\bibitem{liu2020universal}
Xiuying Liu, Shiqi Xia, Ema Jajti{\'c}, Daohong Song, Denghui Li, Liqin Tang,
  Daniel Leykam, Jingjun Xu, Hrvoje Buljan, and Zhigang Chen.
\newblock Universal momentum-to-real-space mapping of topological
  singularities.
\newblock {\em Nat. Commun.}, 11(1):1--8, 2020.

\bibitem{PhysRevA.89.013848}
N.~V. Alexeeva, I.~V. Barashenkov, K.~Rayanov, and S.~Flach.
\newblock Actively coupled optical waveguides.
\newblock {\em Phys. Rev. A}, 89:013848, Jan 2014.

\bibitem{PhysRevA.93.022102}
Stefano Longhi.
\newblock Non-{H}ermitian tight-binding network engineering.
\newblock {\em Phys. Rev. A}, 93:022102, Feb 2016.

\bibitem{PhysRevB.96.161104}
Ajith Ramachandran, Alexei Andreanov, and Sergej Flach.
\newblock Chiral flat bands: {E}xistence, engineering, and stability.
\newblock {\em Phys. Rev. B}, 96:161104, Oct 2017.

\bibitem{PhysRevB.97.045120}
A.~R. Kolovsky, A.~Ramachandran, and S.~Flach.
\newblock Topological flat {W}annier-{S}tark bands.
\newblock {\em Phys. Rev. B}, 97:045120, Jan 2018.

\end{thebibliography}


\end{document}